# Deterministic Computing Mechanism for Perfect Density Classification

[1]**Suryakanta Pal**, [2]**Sudhakar Sahoo** and [3]**Birendra Kumar Nayak**
[1]Silicon Institute of Technology, Silicon Hills, Patia, Bhubaneswar-751024
Email: surya_tuna@yahoo.co.in
[2]Institute of Mathematics and Applications, Andharua, Bhubaneswar -751003
Email: sudhakar.sahoo@gmail.com
[3]P.G. Department of Mathematics, Utkal University, Bhubaneswar-751004
Email: bknatuu@yahoo.co.uk

**Abstract:** The purpose of the present study is to search one-dimensional Cellular Automata (CA) rules which will solve the density classification task (DCT) perfectly. The mathematical analysis of number conserving functions over binary strings of length *n* gives an indication of its corresponding number conserving cellular automata rules (either uniform or non-uniform). The state transition diagrams (STDs) of number conserving CA rules have been analyzed where it has been found that these STDs can generate different DCT solutions. While studying the properties of STDs, an interesting classification of binary strings could be made where equal weight strings form a class and the cardinality of each class is same as the binomial coefficient $^nC_k$; *n* is the length and *k* is the weight of the binary string. Apart from STDs, other deterministic methods have been proposed to obtain the exact solution of DCT. All these exact solutions of DCT using different deterministic methods can be viewed as an improvement over the soft computing techniques used earlier to obtain approximate solutions.
.

## 1. One dimensional density classification task

Density Classification Task (DCT) is one of the simple but most fundamental problem in the theory of Cellular Automata. Consider a one dimensional array of *n*-cells which can take any state from a set of states Q={0, 1, 2,…, *q*-1}. The density $\rho_i$ of state-*i* in a configuration is the ratio of the number of cells with state-*i* and total number of cells. The aim of DCT is to convert all the states into state *i* if $\rho_i > \rho_j$ for all $j \neq i$. For one dimensional two states CA, the standard formulation of DCT defined by Packard (1988)[18] is the following:

An initial configuration converges to a final configuration of all 1s if it has more number of 1s than 0s and to a state of all 0s if it has more 0s than 1s. In case of equality i.e. when the number of 1s is same as the number of 0s, different final configurations have been suggested by different authors. The current research progress and updated sequential results of DCT as studied by various authors are as follows.

1. The first solution to DCT was obtained by Gacs, Kurdyumov and Levin (GKL) (1978)[14]. The CA rule, they had considered, was such that the next state of the cell was the majority amongst (i) the value of the cell on which CA was applied (ii) the value of the cell in immediate neighborhood left (or right) and (iii) the value of the cell which was three units apart to the left (or right). They got an accuracy of 78%.

2. Das, Mitchell and Crutchfield (1994) [8] found solution by applying genetic algorithm which uses particle based computation. But the accuracy was below GKL solution.

3. Land and Belew (1995)[15] proved that there exists no perfect single two-state one-dimensional CA rule under periodic boundary condition, which performs the DCT correctly.

4. M.S. Capcarrere, M. Sipper and M. Tomassini (1996) [4] have proved that there exists a one dimensional two states, radius r=1 CA which can solve DCT perfectly under periodic boundary condition but with a different output specification. They applied CA rule 184 atmost $\lceil N/2 \rceil$ times. If the density of 1's in initial configuration > 0.5(or respectively < 0.5), the iteration leads to a final configuration of one or more blocks of atleast two consecutive 1's (or 0's), interspersed by an alternation of 0s and 1s. If density is exactly 0.5, the final configuration consists of an alternation of 0s and 1s.

5. H. Fuks (1997) [12] showed that a pair of elementary number conserving rules in succession, namely, the traffic rule-184 (applied for $\lfloor (N-2)/2 \rfloor$ times) and the majority rule-232 (applied for $\lfloor (N-1)/2 \rfloor$ times) solve the DCT perfectly.

6. M. Sipper, M. S. Capcarrere and E. Ronald (1998) [21] applied CA rule 184 to a grid of N cells. They showed that any arbitrary initial configuration converges to a configuration of $0^\alpha 1^\beta$ in atmost (N-1) time steps under fixed boundary condition where $\alpha$ and $\beta$ are the number of 0s and 1s in the initial configuration with $\alpha + \beta = N$ and $0 \leq \alpha, \beta \leq N$. If N is odd, the presence of 0(or 1) in middle cell indicates majority of 0s(or 1s). If N is even, the two middle cells provide the decision about density. The presence of 00 indicates majority of 0s, the presence of 11 indicates majority of 1s and presence of 01 indicates uniform density.

7. Chau, Siu and Yan (1999) [3] proved that n-ary multistate density classification is impossible using single one dimensional CA rule but it can be solved using two CA rules in succession.

8. M. S. Capcarrere, M. Sipper (2001) [5] identified two necessary conditions for a perfect density classifier in one dimension which is true only for uniform CA . i.e.
(a)The density of initial configuration must be conserved for all time steps.
(b) A perfect density classifier must exhibit a density of 0.5. i.e. it must be a balanced CA rule.

9. Reynaga and Amthauer (2003) [19] had shown a good performance of DCT when radius of neighborhood is minimum but no 2D CA perfect density classifier exists which classifies properly every configuration in 2D.

10. N.S. Maiti, S. Munshiand and P.P. Chaudhuri (2006) [16] derived the Best Rule Vector (BRV) < 232, 184, 184,…, 184 > for DCT with minimum error using the concept of rule vector graph (RVG) for 3-neighborhood CA and their equivalent rules for k-neighborhood CA (k>3).

11. Wolz and de Oliveira (2008) [23] got an efficacy close to 90% using evolutionary techniques.

12. S. Sahoo, P. Pal Choudhury, A. Pal and B.K. Nayak (2013) [22] used programmable CA for finding 1-D and 2-D DCT solution which classify any initial configuration perfectly under any boundary condition. These rules have been constructed using translation and density preserving properties. The outputs obtained in [21] and [22] are same for all 1-D CA configurations but in [21] the solution is obtained only for fixed boundary condition using a uniform CA rule where as in [22], two sets of $16^4$ non-uniform CA rule vectors are derived for all possible boundary conditions such as null, fixed, periodic, adiabatic, reflexive etc.

13. Naskar and Das (2012) [17] used a non-uniform CA rule which results a homogeneous configuration of $0^N$ or $1^N$ under null boundary condition.

14. P. P. B. de Oliveira (2013) and (2014) [9, 10] had given a detailed survey of DCT with the existing alternate formulations of DCT in the literature along with his personal assessment summarizing some open problems which is left for further investigation.

15. P. P.B. de Oliveira, F. Faria, R. Zanon and R. M. Leite (2015)[11] showed that non-uniform elementary cellular automata with non-local neighbourhoods provide better result than the local neighbourhood for DCT. From computational experiments, they found 4840 non-uniform CA rules in which only 26 CA rules are involved which provides a perfect solution of DCT for a lattice size of 5 cells. But for a lattice of size 7, 9 and 11 cells, the use of evolutionary search techniques could not generate any perfect solution.

16. Markus Redeker (2015) [20] used a set of stochastic CA called the traffic-majority rules to obtain the approximate solution of DCT.

One of the important behaviors of CA rules used in DCT is number conserving which can be well focused through the mathematical modeling of number conserving functions (NCF). To achieve our goal of obtaining number conserving cellular automata rules (NCCA) used in DCT, the concept of number conserving functions over the set of binary strings has been discussed in section-2.

## 2. Number Conserving functions (NCF)

Let $\Sigma = \{0,1,...,\beta-1\}$ where $\beta$ is the base of the number system which is same as number of states in a CA configuration. We define $\Sigma^k = \underbrace{\Sigma \times \Sigma \times .... \times \Sigma}_{k-times} = \{(x_1, x_2, ..., x_k) \mid x_i = 0, 1, ...\text{or } \beta\text{-}1\}$.

Then $\Sigma^k$ be the set of all possible strings over $\Sigma$ of length $k$ where $1 \leq k \leq n$. Let

$\Sigma^{1,2,...,n} = \Sigma^1 \cup \Sigma^2 \cup ....... \cup \Sigma^n = \bigcup_{k=1}^{n} \Sigma^k$. So $\Sigma^{1,2,...,n}$ is the set of all possible strings over $\Sigma$ starting from length 1, length 2 up to length $n$. Let $\Sigma^k_{w_0, w_1,..., w_{\beta-1}}$ be the set of all possible strings of length $k$ where the weights $w_0, w_1, ..., w_{\beta-1}$ are the number of 0's, number of 1's, ..., number of ($\beta-1$)'s respectively such that $w_0 + w_1 + ... + w_{\beta-1} = k$.

**Results: (Relation among sets and their cardinality)**

1. It follows from the definition that $\Sigma^k_{w_0, w_1,..., w_{\beta-1}} \subseteq \Sigma^k \subseteq \Sigma^{1,2,...,n}$.

2. $\left|\Sigma^k_{w_0, w_1,..., w_{\beta-1}}\right| = \dfrac{k!}{w_0! w_1! ... w_{\beta-1}!}$.

Similarly, $\left|\Sigma^k\right| = \beta^k$ and $\left|\Sigma^{1,2,...,n}\right| = \left|\Sigma^1 \cup \Sigma^2 \cup ... \cup \Sigma^n\right| = \left|\Sigma^1\right| + \left|\Sigma^2\right| + ... + \left|\Sigma^k\right| = \beta^1 + \beta^2 + ... + \beta^n$
$= \dfrac{\beta(\beta^n - 1)}{\beta - 1} = \dfrac{\beta^{n+1} - \beta}{\beta - 1}$.

Three functions $f_n : \Sigma^{1,2,...,n} \to \Sigma^{1,2,...,n}$, $g_k : \Sigma^k \to \Sigma^k$ and $h_{k, w_0, w_1,..., w_{\beta-1}} : \Sigma^k_{w_0, w_1,..., w_{\beta-1}} \to \Sigma^k_{w_0, w_1,..., w_{\beta-1}}$ have been defined on three different sets of strings. Following theorems have been utilized to construct the number of NCFs.

**Theorem 1:** The number of functions $f_n : \Sigma^{1,2,...,n} \to \Sigma^{1,2,...,n}$ is $\left(\dfrac{\beta^{n+1} - \beta}{\beta - 1}\right)^{\frac{\beta^{n+1} - \beta}{\beta - 1}}$. Some of these functions preserve the length and weight of the strings.

**Proof**: The number of functions from a set A with $p$ elements to a set B with $q$ elements is $q^p$.

Here A=B=$\Sigma^{1,2,...,n}$. So the number of functions from $\Sigma^{1,2,...,n}$ to $\Sigma^{1,2,...,n}$ is $\left(\dfrac{\beta^{n+1} - \beta}{\beta - 1}\right)^{\frac{\beta^{n+1} - \beta}{\beta - 1}}$.

**Theorem 2:** The number of functions $g_k : \Sigma^k \to \Sigma^k$ is $\left(\beta^k\right)^{\beta^k}$. These functions preserve the length of the strings but may not preserve the weight of the strings.

**Definition: Global vs Local Number conserving functions (NCF)**

A function $f_n^c : \Sigma^{1,2,...,n} \to \Sigma^{1,2,...,n}$ defined by $f_n^c(x) = y$ is global number conserving if it preserves both weight and length of the strings. i.e. if $w(x) = w(y)$ and $l(x) = l(y)$ for all $x \in \Sigma^{1,2,...,n}$. It is a local conserving function if it preserves the weight for all binary strings of a particular length. These number conserving functions (NCF) are of three types depending on three different domains on which they are defined and the formulas of such number of functions are given in theorems 3, 4 and 5.

**Theorem 3:** The number of functions $h_{w_0,w_1,...,w_{\beta-1}} : \Sigma^k_{w_0,w_1,...,w_{\beta-1}} \to \Sigma^k_{w_0,...,w_{\beta-1}}$ is 
$\left( \dfrac{k!}{w_0! w_1! ... w_{\beta-1}!} \right)^{\left( \dfrac{k!}{w_0! w_1! ... w_{\beta-1}!} \right)}$. These functions are NCF as it preserves both weight and length of the strings.

**Theorem 4:** The number of local conserving functions $g^c_k : \Sigma^k \to \Sigma^k$ defined on all strings of length $k$ is $\displaystyle\prod_{w_0+w_1+...+w_{\beta-1}=k} \left( \dfrac{k!}{w_0! w_1! ... w_{\beta-1}!} \right)^{\left( \dfrac{k!}{w_0! w_1! ... w_{\beta-1}!} \right)} = x_k$ (say).

**Proof:** $\Sigma^k$ is the disjoint union taken over $w_0, w_1, ..., w_{\beta-1} \in \{0, 1, ..., \beta-1\}$ such that $w_0 + w_1 + ... + w_{\beta-1} = k$. Number of solutions $w_0, w_1, ..., w_{\beta-1}$ satisfying the equation $w_0 + w_1 + ... + w_{\beta-1} = k$ is same as number of non-negative integral solution of $w_0 + w_1 + ... + w_{\beta-1} = k$ with $0 \le w_0 \le k, 0 \le w_1 \le k, ..., 0 \le w_{\beta-1} \le k$ which is given by $\binom{k+\beta-1}{\beta-1} = \dfrac{(k+\beta-1)!}{(\beta-1)! k!}$. For a particular value of $w_0, w_1, ..., w_{\beta-1}$, number of functions from $\Sigma^k_{w_0,w_1,...,w_{\beta-1}}$ to $\Sigma^k_{w_0,w_1,...,w_{\beta-1}}$ is $\left( \dfrac{k!}{w_0! w_1! ... w_{\beta-1}!} \right)^{\left( \dfrac{k!}{w_0! w_1! ... w_{\beta-1}!} \right)}$. Each possible combination of such function for different value of $w_0, w_1, ..., w_{\beta-1}$ such that $w_0 + w_1 + ... + w_{\beta-1} = k$ gives a complete number conserving function from $\Sigma^k$ to $\Sigma^k$. So the number of NCF $g^c_k : \Sigma^k \to \Sigma^k$ is same as number of such possible combinations which is given by $\displaystyle\prod_{w_0+w_1+...+w_{\beta-1}=k} \left( \dfrac{k!}{w_0! w_1! ... w_{\beta-1}!} \right)^{\left( \dfrac{k!}{w_0! w_1! ... w_{\beta-1}!} \right)} = x_k$. It may be noted that number of terms in the product is $\binom{k+\beta-1}{\beta-1}$.

**Theorem 5:** The number of global number conserving functions $f^c_n : \Sigma^{1,2,...,n} \to \Sigma^{1,2,...,n}$ is $\displaystyle\prod_{i=1}^{n} x_i = x_1 x_2 ... x_n$.

**Proof:** $\Sigma^{1,2,...,n} = \Sigma^1 \cup \Sigma^2 \cup ...... \Sigma^n$. Using theorem:-4, the number of functions $g^c_1 : \Sigma^1 \to \Sigma^1$ which are local number conserving for all strings of length 1 is $\displaystyle\prod_{w_0+...+w_{\beta-1}=1} \left( \dfrac{1!}{w_0! w_1! ... w_{\beta-1}!} \right)^{\left( \dfrac{1!}{w_0! w_1! ... w_{\beta-1}!} \right)} = 1 = x_1$. Similarly, the number of functions $g^c_2 : \Sigma^2 \to \Sigma^2$ which are local number conserving for all strings of length 2 is

$$\prod_{w_0+\ldots+w_{\beta-1}=2}\left(\frac{2!}{w_0!w_1!\ldots w_{\beta-1}!}\right)^{\left(\frac{2!}{w_0!w_1!\ldots w_{\beta-1}!}\right)} = x_2$$ and so on. The number of functions $f_n^c$: $\Sigma^{1,2,\ldots,n} \to \Sigma^{1,2,\ldots,n}$ which are global number conserving for all strings is $x_1 x_2 \ldots x_n = \prod_{i=1}^{n} x_k$.

In particular, $f_n^c(00\ldots0)=00\ldots0$, $f_n^c(11\ldots1)=11\ldots1$ etc.

For $\Sigma=\{0,1\}$, a number conserving function $f_4^c: \Sigma^{1,2,3,4} \to \Sigma^{1,2,3,4}$ is shown in Fig.1.

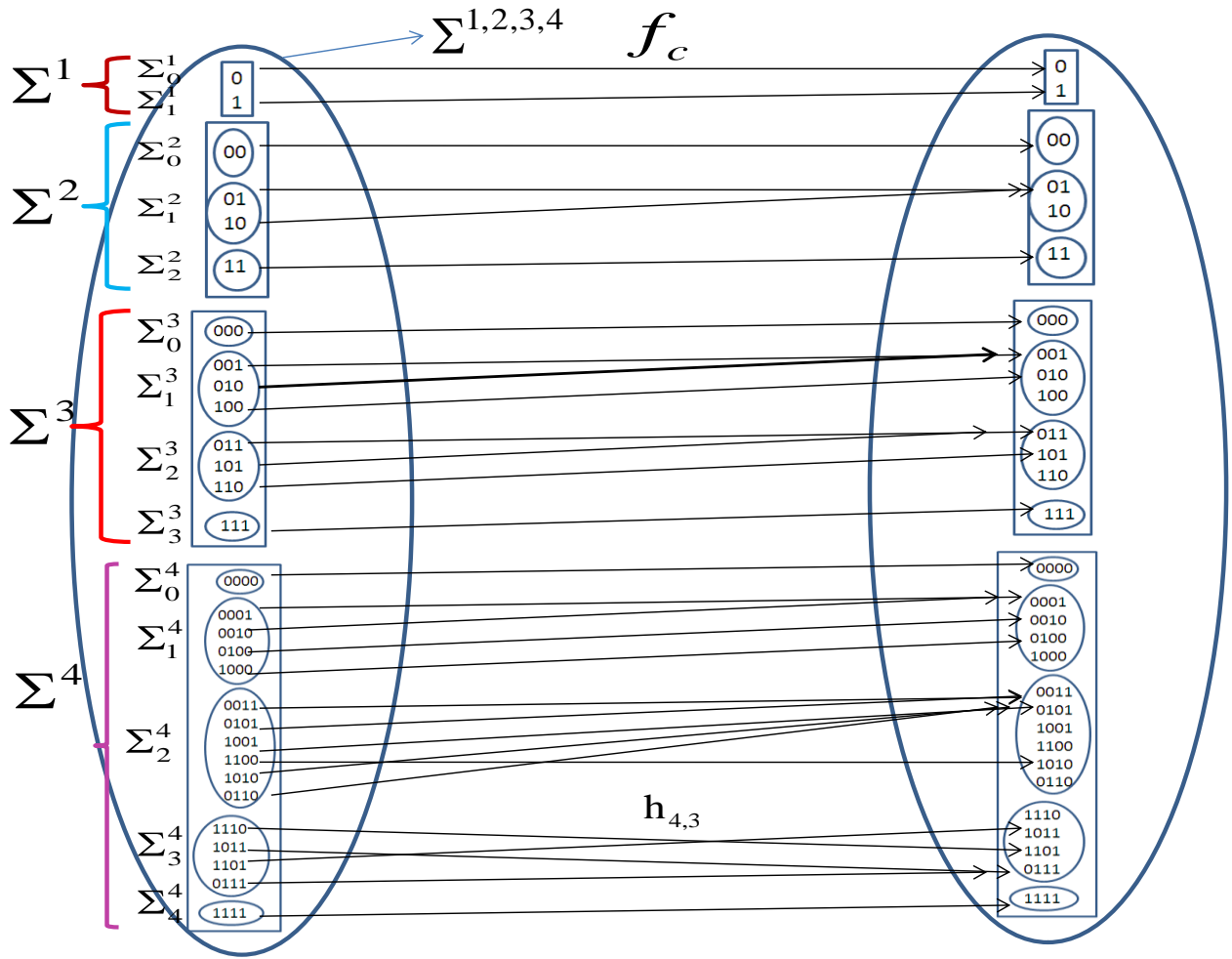

[Fig 1: Shows a number conserving function defined on $\Sigma^{1,2,3,4}$]

We will show the existence of a non-uniform NCCA rule vector with the help of NCFs. Therefore it is required to search the number conserving CA rules from the entire rule space whose cardinality increases exponentially on changing the five different parameters such as number of states, size of neighborhood, length of the CA, guiding rules (uniform or non-uniform) and number of evolutions as reported in [6].

# 3. Non-uniform Number Conserving Cellular Automata Rules

A particular class of CA rules called number conserving cellular automata rules (NCCA) has attracted many researchers to study their dynamical behaviors such as decidability, reversibility, etc. These CA rules are used to generate the same number of 0s and 1s throughout the evolution. A one dimensional CA rule with local update rule f : $Q^n \rightarrow Q$ is number conserving under periodic boundary condition [8] if for all binary strings $x_1 x_2 .... x_L$ of length $L \geq n$,

$$f(x_1, x_2, ..., x_{n-1}, x_n) + f(x_2, x_3, ..., x_n, x_{n+1}) + ......... + f(x_L, x_1, ..., x_{n-1}) = x_1 + x_2 + ... + x_L.$$

From definition, it follows that if a CA rule is NCCA, then $f(x, x, ..., x) = x$. In particular, $f(0, 0, ..., 0) = 0$, $f(1, 1, ..., 1) = 1$ and $\sum_{x_i \in \{0,1\}} f(x_1, x_2, ...., x_n) = 2^{n-1}$.

The necessary and sufficient condition for one-dimensional *q*-states *n*-input CA to be a NCCA derived in [8, 9] is given below.

**Necessary and sufficient condition**

A one dimensional CA rule is number conserving [1, 2, 7, 13] iff for all $(x_1, x_2, .., x_n) \in Q^n$,

$$f(x_1, x_2, ..., x_{n-1}, x_n) = x_1 + \sum_{k=1}^{n-1} \left[ f(\underbrace{0,0,......,0}_{k}, x_2, x_3, ......, x_{n-k+1}) - f(\underbrace{0,0,........,0}_{k}, x_1, x_2, ......, x_{n-k}) \right].$$

Using this definition, it can be proved that out of 256-elementary CA rules, four nontrivial rules, namely, Rule-170, Rule-184, Rule-226 and Rule- 240 are number conserving CA under periodic boundary condition.

The evolution of NCCA rule-184 for an initial configuration 110010100 is shown in Fig. 2 where black cell represents *1* and white cell represents *0*.

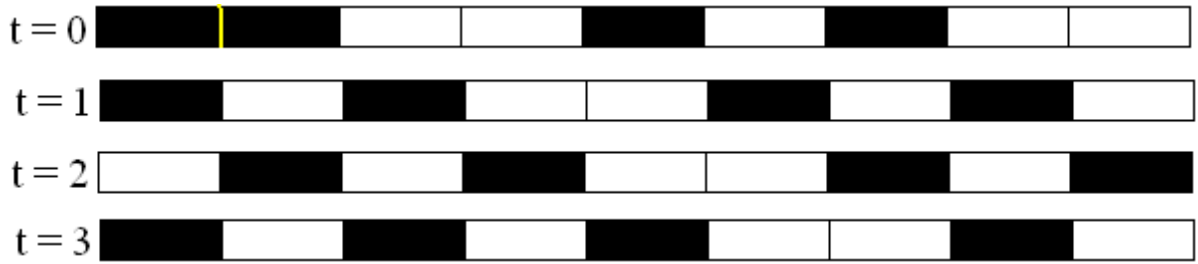

[Fig. 2: Space time diagram for NCCA rule 184]

These elementary NCCA rules are uniform as they are applied to all cells simultaneously. Each uniform elementary NCCA represents a number conserving function. But converse is not true as it is clear from Fig.1. In Fig.1, binary strings 1000 and 0001 are mapped onto 0100 and 0001 under the number conserving function. But there exists no elementary uniform NCCA which represents this mapping. If such a CA exists, then the local function f has the property that f(100)=1 and f(100)=0 which indicates f is not a function. Such situation can be handled if different CA rules are applied for different cells. Thus inadequate number of uniform NCCA motivates us to construct non-uniform elementary NCCA where each cell will be updated according to a CA rule specified in the corresponding rule vector.

To construct non-uniform CA for number conserving function shown in Fig.1, suppose there exists a non-uniform elementary NCCA whose state transition diagram (STD)[5] is shown in Fig.3.

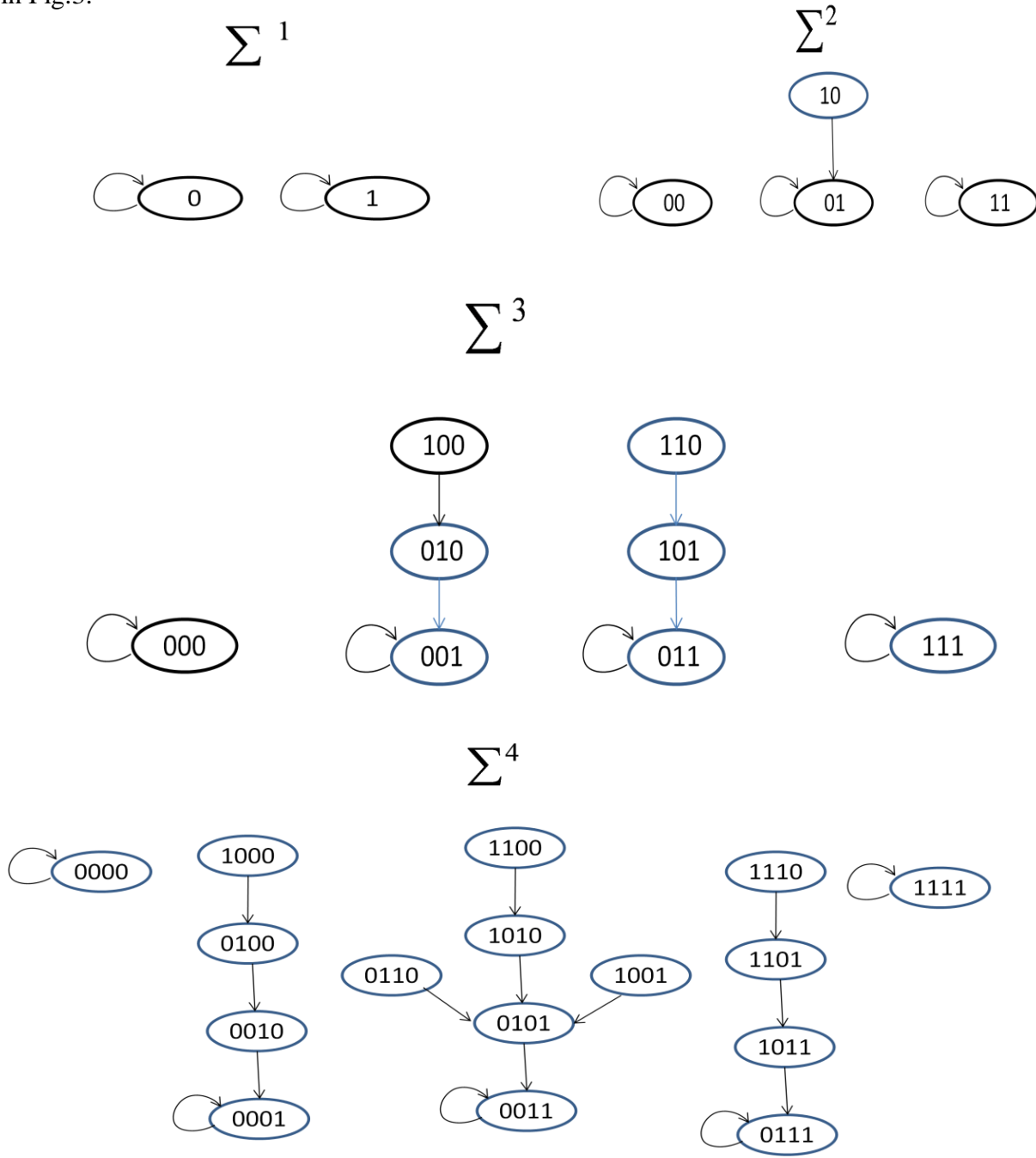

[Fig 3: Shows the state transition diagram of a non-uniform NCCA rule < 136, 184, 184, 252 >]

In Fig.3, a block of 10 in any configuration changes to a block of 01 in next step of evolution. A configuration becomes a stable fixed point attractor of NCCA if it does not contain a block of 10. So CA rule-184 is applied in all cells except at the boundaries as CA rule-184 has such dynamic property. Since the periodic boundary condition is used and the density of 1's will be preserved throughout the evolution so we apply CA rule-136 for first cell and rule-252 for last cell. So the corresponding CA rule vector is < 136, 184, 184, 252 >. Generalization of this leads to the following theorem.

**Theorem:-6** A non-uniform elementary number conserving CA rule vector for any CA configuration of any length under periodic boundary condition is < 136, 184, …, 184, 252 >. Similarly, another dynamically equivalent non-uniform NCCA is <238, 226, 226, …, 226, 192 >.

Every number conserving CA (either uniform or non-uniform) are examples of number conserving functions. But all number conserving functions cannot be represented by non-uniform elementary NCCA. It is because if for a number conserving function, $f_5^c$ (10110)=01011 and $f_5^c$ (00111)=01101, then for a particular CA rule applied in third cell, f(011)=0 and f(011)=1. It shows that f is not a function. Thus a NCF possessing such type of property can not represent a NCCA. As CA rules are necessarily functions so a systematic procedure should be investigated to identify those number conserving functions which can represent a CA rule.

# 4. Deterministic methods for solving DCT

The algorithm for solving 1-D DCT problem works in two phases, namely, a preprocessing phase and a decision phase [22]. In preprocessing phase, a non-uniform CA acts on an initial configuration of length N for (N-1) times to get a particular type of configuration in which the block of all 1's are either in the left or in the right. In decision phase, the density of the string can be ascertained by observing either the middle or the two consecutive middle bits depending on odd and even length of the initial configuration. In this paper, with reference to algorithm given in [22], we have adopted two different approaches to find the exact solution of DCT.

Method I: Changing the criteria in decision phase and keeping the preprocessing phase unchanged.

Method II: Changing the logic of preprocessing phase and then taking decision without changing the decision algorithm in the decision phase.

## 4.1 Method-I

In this method, we have changed the decision phase but the preprocessing phase remains unchanged. In preprocessing phase, we separate the 0's from 1's by forming a string of the form $1^x0^y$ or $0^x1^y$. After this, we decrease the number of 1's on applying rule-192 and rule-13 uniformly for ($\lfloor N/2 \rfloor$-1) times to reach a configuration looking at which decision can be made.

**Preprocessing Phase**

Consider a binary string $x_1x_2x_3………x_N$ of length N. We will find a CA rule vector which translates all 1's of $x_1x_2x_3………x_N$ to the left and all 0's to the right under any boundary condition in such a way that density of 0's and 1's are preserved throughout evolution process. It can be achieved by applying the rule vector $< 238, 226, 226, \ldots, 226, 192 >$ for (N-1) times. After applying it, the given string reduces to the form $1^x 0^y$. Then we apply the rule-136 for ($\lfloor N/2 \rfloor -1$) times. In this method the total number of iterations required is ($N + \lfloor N/2 \rfloor - 2$).

## Decision phase

(i) If the first two bits are 11, then the string is dense in 1.
(ii) If the first two bits are 00, then the string is dense in 0.
(iii) If the first two bits are 10 and
        Case 1: if the input string is of odd length then the string is dense in 0.
        Case 2: if the input string is of even length then the string is of uniform density.

**Illustration 1:** Consider a binary string of length N=9.

| | | | | | | | | | |
|---|---|---|---|---|---|---|---|---|---|
| Initial configuration: | 1 | 0 | 0 | 1 | 1 | 0 | 1 | 1 | 0 |
| For t=1: | 1 | 0 | 1 | 0 | 1 | 1 | 0 | 1 | 0 |
| For t=2: | 1 | 1 | 0 | 1 | 0 | 1 | 1 | 0 | 0 |
| For t=3: | 1 | 1 | 1 | 0 | 1 | 0 | 1 | 0 | 0 |
| For t=4: | 1 | 1 | 1 | 1 | 0 | 1 | 0 | 0 | 0 |
| For t=5: | 1 | 1 | 1 | 1 | 1 | 0 | 0 | 0 | 0 |
| For t=6: | 1 | 1 | 1 | 1 | 1 | 0 | 0 | 0 | 0 |
| For t=7: | 1 | 1 | 1 | 1 | 1 | 0 | 0 | 0 | 0 |
| For t=8: | 1 | 1 | 1 | 1 | 1 | 0 | 0 | 0 | 0 |

-----------------------------------------

| | | | | | | | | | |
|---|---|---|---|---|---|---|---|---|---|
| For t=9: | 1 | 1 | 1 | 1 | 0 | 0 | 0 | 0 | 0 |
| For t=10: | 1 | 1 | 1 | 0 | 0 | 0 | 0 | 0 | 0 |
| For t=11: | 1 | 1 | 0 | 0 | 0 | 0 | 0 | 0 | 0 |

In the last iteration the first two bit is 11. So, the string is dense in 1.

In similar process, we can apply the rule vector $<136, 184, 184, \ldots, 184, 252>$ to get a string of the form $0^x 1^y$ and then apply rule-192 to decrease the number of 1's. In this case if the last two bits are 11, then the string is dense in 1. If the last two bits are 00, then the string is dense in 0. If last two bits are 01, then the string is dense in 0's for odd length and is of uniform density for even length.

**Illustration 2:** Consider the same binary string of illustration 1.

| | | | | | | | | | |
|---|---|---|---|---|---|---|---|---|---|
| Initial configuration: | 1 | 0 | 0 | 1 | 1 | 0 | 1 | 1 | 0 |
| For t=1: | 0 | 1 | 0 | 1 | 0 | 1 | 1 | 0 | 1 |
| For t=2: | 0 | 0 | 1 | 0 | 1 | 1 | 0 | 1 | 1 |
| For t=3: | 0 | 0 | 0 | 1 | 1 | 0 | 1 | 1 | 1 |
| For t=4: | 0 | 0 | 0 | 1 | 0 | 1 | 1 | 1 | 1 |
| For t=5: | 0 | 0 | 0 | 0 | 1 | 1 | 1 | 1 | 1 |
| For t=6: | 0 | 0 | 0 | 0 | 1 | 1 | 1 | 1 | 1 |
| For t=7: | 0 | 0 | 0 | 0 | 1 | 1 | 1 | 1 | 1 |
| For t=8: | 0 | 0 | 0 | 0 | 1 | 1 | 1 | 1 | 1 |

-------------------------------------------------
- - -

| For t=9:  | 0 0 0 0 0 1 1 1 1 |
| For t=10: | 0 0 0 0 0 0 1 1 1 |
| For t=11: | 0 0 0 0 0 0 0 1 1 |

In the last iteration the first two bit is 11. So the string is dense in 1.

In a reverse approach after converting the initial configuration to a form $1^x0^y$ or $0^x1^y$, one can increase the number of 1's on applying the rule-252 and rule-238 respectively to take the decision.

Instead of translating the 1's to either end it is possible to translate the 1's to the middle from both ends to convert it into a string of the form $0^x1^y0^z$. This can be achieved by applying the rule vector is <136, 184, 184, …, 184, 252, 238, 226, 226, …, 226, 192> for (N-1) times. Then decision can be taken about the density of the string.

Although the above method generates new CA rules for DCT but number of such rule vectors are very few. These inadequate numbers of rules have diverted the attention of some researchers towards heuristic methods to get the approximate solution. But we are interested for exact solution of DCT. To achieve our goal we have used these NCCA rules to draw their STDs which are the basis for the existence of other CA rules to solve the DCT problem perfectly.

## 4.2 Method-II
In this method we have changed the logic of preprocessing phase using STD and array generated from NCCA rules. Then decision is made using the same procedure as adopted in [22].

### 4.2.1 DCT solution using State Transition Diagram
The dynamic behavior of cellular automaton evolution can be completely described using state transition diagram (STD) which is a directed graph in which configurations are represented by nodes and two configurations are connected by an edge representing an evolution. The STD generated by applying the non-uniform rule vector < 238, 226, 226, 226, 192 > to all possible 32 binary strings of length 5 is shown in Fig-4.

**Properties of STD**
1. STD gives a partition of set of all binary strings of length $n$ into $(n+1)$ mutually disjoint equivalence classes where the strings in same class have same weight(weight of a string is equal to number of 1's in it). Suppose class-k, k=0, 1, 2, …, n contains all possible binary strings of length $n$ having weight $k$.
2. Any attractor of the rule vector <238, 226, 226, 226, 192> is of the form $1^x0^y$ where $0 \leq x \leq 5$, $0 \leq y \leq 5$ and x+y=5. Attractor of class-k of all binary strings of length $n$ is $2^n - 2^{n-k}$, k =0, 1, 2, 3, ...., n.

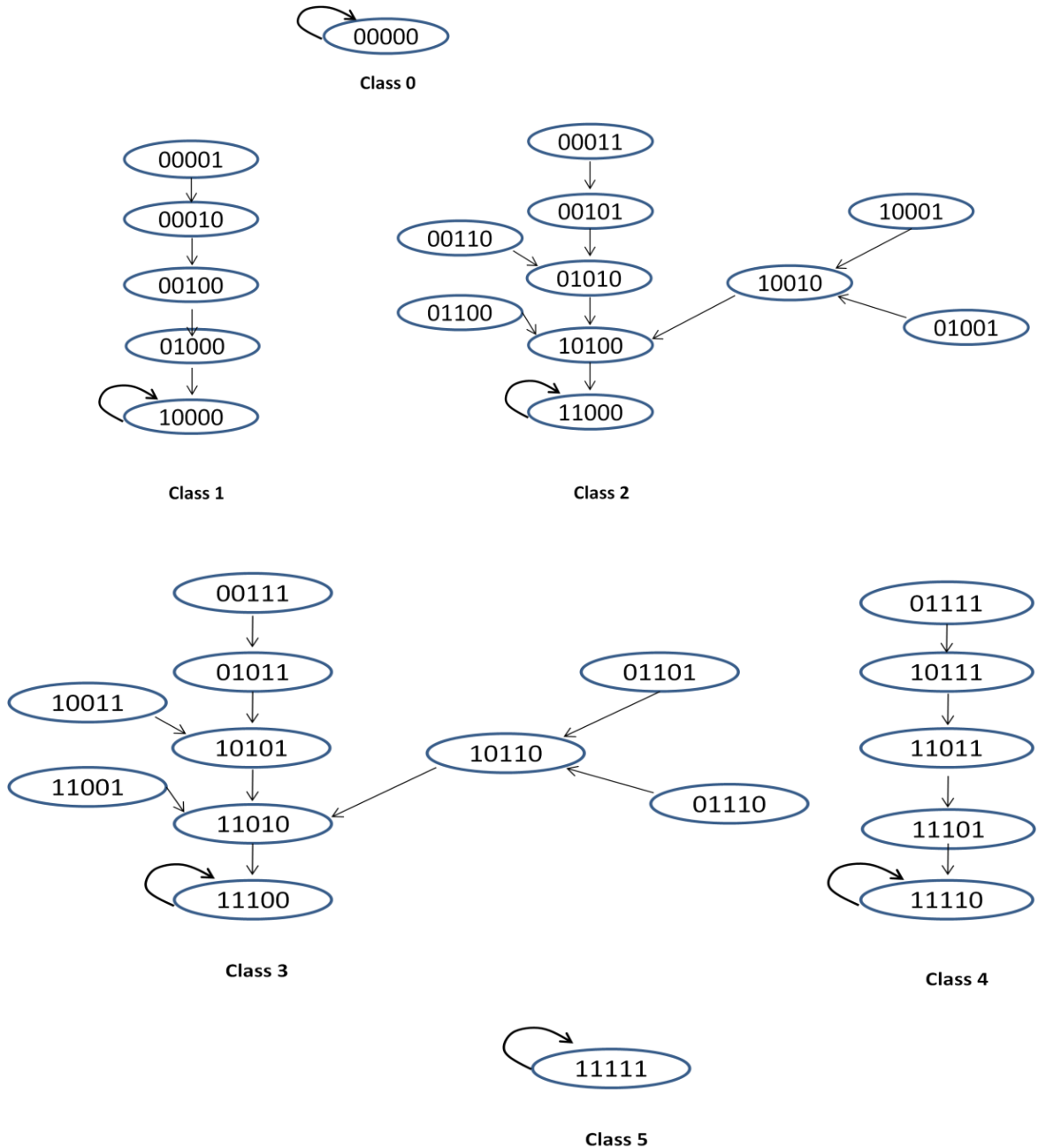

[Fig 4: Shows STD of rule vector $< 238, 226, 226, 226, 192 >$]

**Proof:** Let $x^t = [x_1 \ x_2 \ x_3 \ x_4 \ x_5]$ be attractor of the 5 cell CA.
Since $f_{238}(x_{i-1}, x_i, x_{i+1}) = x_i \oplus x_{i+1} \oplus x_i x_{i+1}$, $f_{226}(x_{i-1}, x_i, x_{i+1}) = x_{i+1} \oplus x_{i-1} x_i \oplus x_i x_{i+1}$ and $f_{226}(x_{i-1}, x_i, x_{i+1}) = x_{i-1} x_i$ so next state is given by
$$x^{t+1} = [x_1 \oplus x_2 \oplus x_1 x_2 \quad x_3 \oplus x_1 x_2 \oplus x_2 x_3 \quad x_4 \oplus x_2 x_3 \oplus x_3 x_4 \quad x_5 \oplus x_3 x_4 \oplus x_4 x_5 \quad x_4 x_5]$$
For single cycle attractors of CA, $x^{t+1} = x^t$.

$\Rightarrow x_1 \oplus x_2 \oplus x_1 x_2 = x_1$ .........................(1)

$x_3 \oplus x_1 x_2 \oplus x_2 x_3 = x_2$ .........................(2)

$x_4 \oplus x_2 x_3 \oplus x_3 x_4 = x_3$ .........................(3)

$x_5 \oplus x_3 x_4 \oplus x_4 x_5 = x_4$ .........................(4)

$x_4 x_5 = x_5$ ...........................................(5)

From equation (4) and (5), we get $x_3 x_4 = x_4$ ................................(6)

From equation (3) and (6), we get $x_2 x_3 = x_3$ ................................(7)

From equation (2) and (7), we get $x_1 x_2 = x_2$ ................................(8)

From above equations, we get $x_i x_{i+1} = x_{i+1}$ for i=1, 2, 3 and 4. Thus if $x_i = 0$, then $x_{i+1} = 0$.

It means if any bit of the attractor is zero then all bits after it up to last bit are zero.

If $x_i = 1$, then $x_{i+1} = 0$ or 1. So the attractors are $(00000)_2=0$, $(10000)_2=16$, $(11000)_2=24$, $(11100)_2=28$, $(11110)_2=30$, $(11111)_2=31$.

In general, any attractor of the rule vector < 238, 226, 226,..., 226, 192 > is of the form $1^x 0^y$ where $0 \leq x \leq n$, $0 \leq y \leq n$ and x+y=n.

The set of attractors is

$A = \{0,\ 2^{n-1},\ 2^{n-1}+2^{n-2},\ 2^{n-1}+2^{n-2}+2^{n-3},\ \ldots,\ 2^{n-1}+2^{n-2}+2^{n-3}+\ldots+1\}$.

$= \{0,\ 2^n - 2^{n-1},\ 2^n - 2^{n-2},\ 2^n - 2^{n-3},\ \ldots,\ 2^n - 1\}$

Thus attractor of class-k of all binary strings of length $n$ is $2^n - 2^{n-k}$, k =0, 1, 2, 3, …., n.

**3.** Each of the *01* block of any string gets converted to a block of *10* in next evolution. So a string is stable (i.e. reaches an attractor) only when there is no block of *01*.

**4**. The cardinality of class-k is the binomial coefficient C(n, k)= $\dfrac{n!}{k!(n-k)!}$.

**5.** The elements of class-1 are $e_1 = (1,0,\ldots,0,0)$, $e_2 = (0,1,\ldots,0,0,0), \ldots, e_n = (0,0,\ldots,0,0,1)$ which are the basis elements of $R^n$. Elements of class-(n-1) are $2^n - 2$ and $2^n - 2 - (2^0 + 2^1 + 2^2 + \ldots + 2^m) = 2^n - 2 - (2^{m+1} - 1) = 2^n - 2^{m+1} - 3$ where $0 \leq m \leq n-2$.

**6.** The complement of class-k is the class-(n-k) where k=0, 1,2,………..,n but $k \neq \dfrac{n}{2}$. If $k = \dfrac{n}{2}$, then class-k contains some strings and their complements.

**7**. The class-0, class-1, class-2, …, class-$\lfloor n/2 \rfloor$ are dense in 0's and the remaining classes are dense in 1's. But if *n* is even, then class-(*n*/2) contains the strings of uniform density.

From these STDs, density of an arbitrary string can be obtained using the following algorithm.

## Algorithm-1
*Input:* An arbitrary binary string of length *n*.
*Output:* Decision about density.
1. Take an *n*-bit string.
2. Identify the string in the STD and traverse the successor states till the attractor is reached.
3. Apply the decision phase logic given in [22] to remark about its density.

If $n$ is odd then STDs of class-0, class-1,…, class-$\lfloor n/2 \rfloor$ can be interconnected through the attractors of each class to get an attractor with all 0's (Fig.5(a)). This can be achieved either using the uniform CA rule 0 or using the non-uniform rule vector < 242, 46, 46, …, 46, 252 >. These rule vectors are not unique. It is because STDs of class-0, class-1, …, class-$\lfloor n/2 \rfloor$ are dense in 0. So middle bit of the attractor of these classes must be 0. Since we have already translated all 1's to the left so the attractor in (N-1)-th iteration must be of the form $1^x 0^y$. To get a state of all 0's as an attractor in next step, we can construct different non-uniform CA rules by assigning a particular value either 0 or 1 to different possible neighborhoods except the impossible neighborhoods like 001, 101 which can be assigned by either 0 or 1(i.e. don't care $d$) as shown in table-1. The same procedure can be applied to the remaining classes to get another attractor with all 1's (Fig. 5(b)) which can be achieved by using either the uniform CA rule 255 or using the non-uniform rule vector < 136, 209, 209, …, 209, 162 >. They will produce two basins of attractors where former is dense in 0's and the latter is dense in 1's.

| Cells | 111 | 110 | 101 | 100 | 011 | 010 | 001 | 000 | CA rules |
|---|---|---|---|---|---|---|---|---|---|
| For $x_1$ | d | d | d | d | 0 | 0 | d | 0 | 0, 2, 16, 18, 32, 34, 48, 50, 64, 66, 80, 82, 96, 98, 112, 114, 128, 130, 144, 146, 160, 162, 176, 178, 192, 194, 208, 210, 224, 226, 240, 242. |
| For $x_2$ | 0 | 0 | d | 0 | d | d | d | 0 | 0, 2, 4, 6, 8, 10, 12, 14, 32, 34, 36, 38, 40, 42, 44, 46. |
| For $x_3$ | 0 | 0 | d | 0 | d | d | d | 0 | 0, 2, 4, 6, 8, 10, 12, 14, 32, 34, 36, 38, 40, 42, 44, 46. |
| . | . | . | . | . | . | . | . | . | . |
| . | . | . | . | . | . | . | . | . | . |
| . | . | . | . | . | . | . | . | . | . |
| For $x_{(\lceil N/2 \rceil - 2)}$ | 0 | 0 | d | 0 | d | d | d | 0 | 0, 2, 4, 6, 8, 10, 12, 14, 32, 34, 36, 38, 40, 42, 44, 46. |
| For $x_{(\lceil N/2 \rceil - 1)}$ | d | 0 | d | 0 | d | d | d | 0 | 0, 2, 4, 6, 8, 10, 12, 14, 32, 34, 36, 38, 40, 42, 44, 46, 128, 130, 132, 134, 136, 138, 140, 142, 160, 162, 164, 166, 168, 170, 172, 174. |
| For $x_{(\lceil N/2 \rceil)}$ | d | d | d | 0 | d | d | d | 0 | 0, 2, 4, 6, 8, 10, 12, 14, 32, 34, 36, 38, 40, 42, 44, 46, 64, 66, 68, 70, 72, 74, 76, 78, 96, 98, 100, 102, 104, 106, 108, 110, 128, 130, 132, 134, 136, 138, 140, 142, 160, 162, 164, 166, 168, 170, 172, 174. |
| For $x_{(\lceil N/2 \rceil + 1)}$ | d | d | d | d | d | d | d | 0 | All even numbered rules up to 254 |
| . . . | . . . | . . . | . . . | . . . | . . . | . . . | . . . | . . . | . . . |
| For $x_{N-1}$ | d | d | d | d | d | d | d | 0 | All even numbered rules up to 254. |
| For $x_N$ | d | d | d | d | d | d | d | 0 | All rules which are multiple of 4 upto 252 |

Table 1: Shows the CA rules which acts on all attractors of class-0, class-1, …., class-$\lfloor n/2 \rfloor$ to get a single attractor of all 0's.

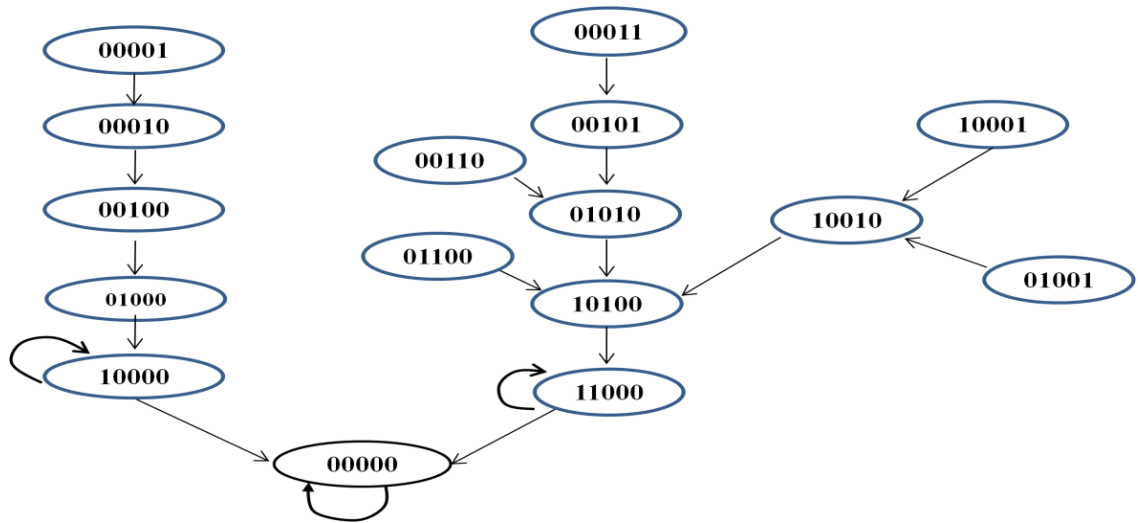

(a)

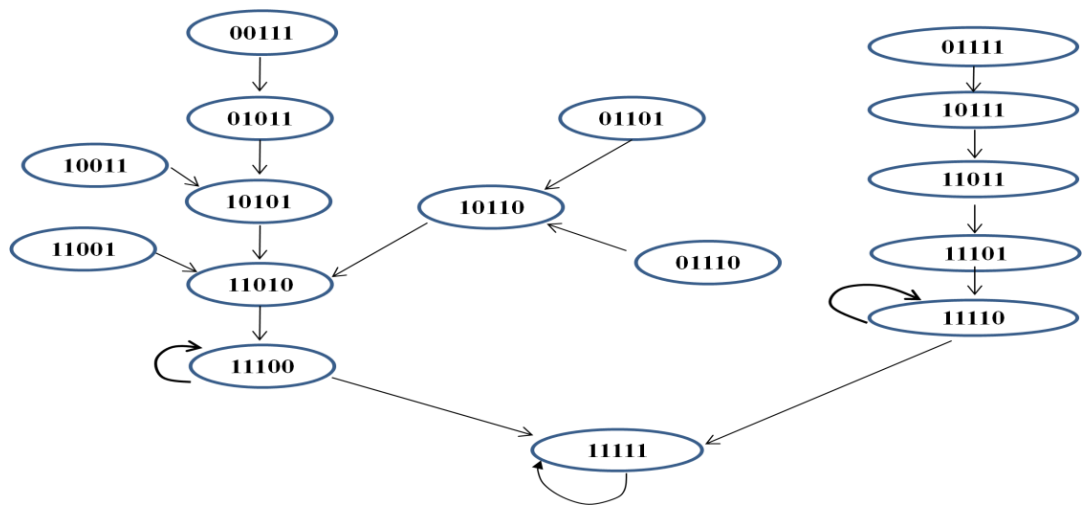

(b)

[Fig 5: (a) shows all strings are dense in 0 and (b) shows all strings are dense in 1.]

If *n* is even then the similar grouping produces three basin of attractors using the classes 0 to (n/2)-1, class(n/2) +1 to n and class (n/2) giving an exact solution of DCT as dense in 0's, dense in 1's and equal density respectively.

The intra-class and interclass connection of different classes of STDs can be done in several ways to generate different trees like structure which lead to a solution of DCT. Whether there exists a CA rule vector for each such STD is a challenging task. It is because drawing STDs from a given CA is easy but the reverse is difficult even if we relax the different parameters of CA. It can be observed that number conserving property is not the only criteria of CA rules for solving the DCT problem. In a similar way the rule vector < 136, 184, 184,…, 184, 252 > can also generates enumerable number of STDs for the solution of DCT.

### 4.2.2 DCT solution using 1-D array

The rule vector < 238, 226, 226,…, 226, 192 > generates a STD which is a tree-like structure except attractor which forms a loop. For this STD, we can find an upper triangular adjacency matrix of size $(2^n-1) \times (2^n-1)$ where the ordered pair of attractors corresponds to diagonal entries of 1's with the diagonal entries $a_{ii} = \begin{cases} 1 & \text{if } i=2^n - 2^{n-k}, k=0, 1,...., n-1 \\ 0, & \text{otherwise} \end{cases}$

The number corresponding to a column of 0's is a leaf node. To get rid of space complexity of adjacency matrix for large value of *n*, we can represent the same thing using the successor array.

For n=3, the successor array is

| Successors | 0 | 2 | 4 | 5 | 4 | 6 | 6 | 7 |

For n=4, the successor array is

| Successors | 0 | 2 | 4 | 5 | 8 | 10 | 10 | 11 | 8 | 10 | 12 | 13 | 12 | 14 | 14 | 15 |

For n=5, the successor array is

| Successors | 0 | 2 | 4 | 5 | 8 | 10 | 10 | 11 | 16 | 18 | 20 | 21 | 20 | 22 | 22 | 23 |

| continued | 16 | 18 | 20 | 21 | 24 | 26 | 26 | 27 | 24 | 26 | 28 | 29 | 28 | 30 | 30 | 31 |

Let $s_k^n$ =decimal value of successor of node *k* (in decimal) having *n*-bits in its binary form.
The values of successors for n-bits can be obtained from that of (*n*-1)-bits using the relationship

$$s_k^n = \begin{cases} s_k^{n-1} & \text{if } 0 \leq k \leq 2^{n-2} - 1 \\ s_k^{n-1} + 2^{n-2} & \text{if } 2^{n-2} \leq k \leq 2^{n-1} - 1 \\ s_{k-2^{n-1}}^{n-1} + 2^{n-1} & \text{if } 2^{n-1} \leq k \leq 2^n - 1 \end{cases}$$

### Algorithm:-2
*Input*: An arbitrary binary string of length *n*.
*Output*: Decision about density.
(i) Find the decimal value of the input string.

(ii) Find the corresponding column in successor array.
(iii) Find the value present in the corresponding column in successor array.
(iv) Repeat step (ii) and step (iii) till the column number = value present in that column=d (say).
(v) If $n$ is even and d= 0, $2^n - 2^{n-1}$, $2^n - 2^{n-2}$, $2^n - 2^{n-3}$,.......... or $2^n - 2^{n-[(n/2)-1]}$, then the input string is dense in 0. But for d= $2^n - 2^{n/2}$, the string is of uniform density. Otherwise, the string is dense in 1. If $n$ is odd and d $=0$, $2^n - 2^{n-1}$, $2^n - 2^{n-2}$, $2^n - 2^{n-3}$,......... or $2^n - 2^{n-\lfloor n/2 \rfloor}$, then string is dense in 0. Otherwise, it is dense in 1.

**Illustration 3**: For $n=4$ and $(0111)_2=7$, we have $7 \rightarrow 11 \rightarrow 13 \rightarrow 14 \rightarrow (14, 14)$. So d=14= $2^4 - 2 =$ $2^n - 2^{n-3}$. So the string 0111 is dense in 1.

# 5. Conclusion and Future research Directions

This paper gives a brief description about number conserving rules used in solution of DCT which lead to perfect classification. We have shown that there are several methods which can be implemented to obtain correct classification the density. In particular, our analysis on state transition diagrams (STDs) of CA rules can generate different DCT solutions. As an example, the STDs of class-0, class-1, …, class-$\lfloor n/2 \rfloor$ can be interconnected by rule matrix

$$R_1 = \begin{pmatrix} 238 & 226 & 226 & ... & 226 & 192 \\ 238 & 226 & 226 & ... & 226 & 192 \\ \vdots & \vdots & \vdots & & \vdots & \vdots \\ 238 & 226 & 226 & ... & 226 & 192 \\ 242 & 46 & 46 & ... & 46 & 252 \end{pmatrix}_{n \times n}$$

where the rule vector of the $i$-th row generates the $i$-th evolution and the last rule vector generates a string of 0's. It is observed that in $n$-th iteration the attractors of above classes are connected to get a single attractor of all 0's. Similarly, rule matrix

$$R_2 = \begin{pmatrix} 238 & 226 & 226 & ... & 226 & 192 \\ 238 & 226 & 226 & ... & 226 & 192 \\ \vdots & \vdots & \vdots & & \vdots & \vdots \\ 238 & 226 & 226 & ... & 226 & 192 \\ 136 & 209 & 209 & ... & 209 & 162 \end{pmatrix}_{n \times n}$$

can be applied to the remaining STDs to get another attractor of all 1's. These rule vectors are not unique and many such rule matrices can be constructed. Thus the STD's of the above rule matrices are mathematical models which can generate different DCT solutions though their CA rule vectors needs to be calculated.

So, our next effort is to search rule vectors (perhaps using evolutionary techniques) for these different interconnected networks over the space of all CA rules by relaxing different parameters of CA. We will also extend this 1-D work to two-dimensional density classification task.